\documentclass[pre,aps,preprint,11pt,amsmath,amssymb,floatfix,superscriptaddress]{revtex4-1}

\usepackage{graphicx}
\usepackage{dcolumn}
\usepackage{bm}

\usepackage[usenames]{color}

\begin{document}

\title{Reconciling transport models across scales: the role of volume exclusion} 

\author{P. R. Taylor}
\email[]{paul.taylor@maths.ox.ac.uk}
\affiliation{Mathematical Institute, University of Oxford, Woodstock Road, Oxford, OX2 6GG, UK}

\author{C. A. Yates}
\affiliation{Department of Mathematical Sciences, University of Bath, Claverton Down, Bath, BA2 7AY, UK}

\author{M. J. Simpson}
\affiliation{Mathematical Sciences, Queensland University of Technology, G.P.O. Box 2434, Queensland 4001, Australia}

\author{R. E. Baker}
\affiliation{Mathematical Institute, University of Oxford, Woodstock Road, Oxford, OX2 6GG, UK}

\date{\today}

\begin{abstract}
Diffusive transport is a universal phenomenon, throughout both biological and physical sciences, and models of diffusion are routinely used to interrogate diffusion-driven processes. However, most models neglect to take into account the role of volume exclusion, which can significantly alter diffusive transport, particularly within biological systems where the diffusing particles might occupy a significant fraction of the available space. In this work we use a random walk approach to provide a means to reconcile models that incorporate crowding effects on different spatial scales. Our work demonstrates that coarse-grained models incorporating simplified descriptions of excluded volume can be used in many circumstances, but that care must be taken in pushing the coarse-graining process too far. 
\end{abstract}


\maketitle 

% ----------------------------------------------------------------------------------------------------

\section{Introduction}

Throughout the physical and biological sciences, diffusive transport is ubiquitous, and it takes places on a wide range of spatial and temporal scales. For example, in biology diffusion is a key transport process that regulates events and interactions on levels ranging from those describing the behaviours of ions and subcellular macromolecules, to those of cells, tissues and organisms~\cite{Chowdhury:2005:POT}. Less well understood, however, is how volume-exclusion-driven crowding impacts upon these diffusive transport processes, despite the inherent fact that all diffusing particles exclude other particles from occupying the same region in space~\cite{Dix:2008:CEO,Dlugosz:2011:DIC}.

The majority of models of diffusive processes neglect to take into account excluded volume effects. The predominant models describing diffusive transport over a range of spatial scales, and with varying excluded volume fractions, are `diffusion' partial differential equation (PDE) models with a constant diffusion coefficient~\cite{Lander:2002:DMG,Maini:2004:TWW,Khain:2008:GCH}, and random-walk-based models of point particles~\cite{Chandrasekhar:1943:SPP,Othmer:1997:ABC}, both of which entirely neglect the impact of volume exclusion. Other models include phenomenological descriptions of volume exclusion effects by imposing that, for example, the diffusion coefficients of PDE models or the `jump rates' associated with random walk models depend locally on the particle density~\cite{Anguige:2009:ODM,Black:2012:SFE,Painter:2003:MMI,Thompson:2012:MCA,Turner:2004:FDC}. However, these phenomenological descriptions are usually chosen on an \textit{ad hoc} basis, and the ramifications of choosing one description over another rarely explored in detail.

In this work we employ the framework of a lattice-based random walk to explore how volume exclusion may be taken into account at different spatial scales, and describe how to define the jump rates of the random walkers so as to provide a consistent description of the effects of volume exclusion across spatial scales. An additional advantage of our approach is that it provides for significant time savings in the computational simulation of volume excluding individual-particle-level models of diffusive transport.

% ----------------------------------------------------------------------------------------------------

\section{Modelling diffusive transport on different scales} \label{sec:models}

For simplicity, we consider a one-dimensional, lattice-based random walk model of diffusive transport in which the motile particles have length $h$. We work with the domain $x\in[0,L]$, where $L=Nh$ for some $N\in\mathbb{N}$, so that the domain can hold at most $N$ particles. We impose a uniform lattice consisting of $K$ compartments on the domain, so that the length of each compartment is $L/K$, and we work only with choices of $K$ for which $m=L/(Kh)$ is a positive integer. This means that $N=Km$ and at most $m$ particles fit into each of the $K$ compartments~\cite{Anguige:2009:ODM,Baker:2010:FMM,Khain:2014:SPO,Simpson:2009:MSS}. We model diffusion as a series of jumps between compartments, and impose zero-flux boundary conditions on the domain (so that particle number is conserved). 

% ----------------------------------------------------------------------------------------------------

\begin{figure}[h]
\includegraphics{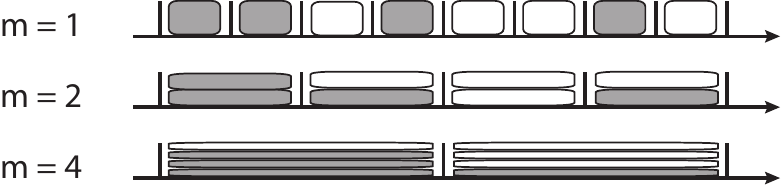}
\centering
\caption{Representation of particle positions within compartments of different capacity. Shaded cells represent particles, and white cells represent unused capacity. As $m$ increases, the spatial resolution coarsens, and this work reconciles the descriptions at these different scales.}
\label{fig:CroppedCoarseningLatticeDiagram}
\end{figure}

% ----------------------------------------------------------------------------------------------------

We move between different levels of spatial resolution by varying the compartment capacity, $m$: smaller values of $m$ resolve changes in particle density on a finer spatial scale than larger values. The two limiting cases are: full exclusion, $m=1$, so that compartments contain at most one particle (and the position of the particle is fixed); and no exclusion, $m\rightarrow\infty$, so there is no limit on the number of particles per compartment. In this work, we shall term the $m=1$ case `accurate', in the sense that no assumptions are made on the positions of particles with each compartment.

The common, phenomenological approach taken in the literature is to define the jump rates between compartments as
\begin{equation}
\label{eq:TranisitionProbabilities}
T_j^\pm=\frac{D}{m^2h^2}\left[1-f^{(m)}\left(n^{(m)}_{j\pm1}\right)\right], \quad j=1,\ldots,K,
\end{equation}
where $n^{(m)}_j$ is the number of particles in compartment $j$ when each compartment has capacity $m$~\cite{Painter:2003:VFQS}. The function $f^{(m)}$ describes the effects of volume exclusion, effectively specifying the proportion of jumps that `fail' due to crowding~\cite{footnote1}. The scaling of the jump rate with the square of the compartment size ($mh$) can be justified from mean first passage time approaches~\cite{Redner:2001:GFP}. Effects such as adhesion are often included by assuming the $T_j^\pm$ to also be a function of $n_j^{(m)}$ and $n_{j\pm1}^{(m)}$~\cite{Anguige:2009:ODM,Painter:2003:MMI}.

Typically, these models on different scales are interrogated either by: (i) using repeated simulation of the random walk models to estimate summary statistics of interest; (ii) deriving and solving ordinary differential equations (ODEs) for the expected particle number per compartment; or (iii) deriving PDE models in the limit $h\rightarrow0$ and using standard analytical and numerical techniques for PDEs to explore model behaviours. However, to the best of our knowledge, there has been little exploration of the effects of choosing different functional forms for $f^{(m)}$ on various summary statistics of the random walk models as the compartment capacity, $m$, varies. As such, one of the aims of this work is to understand how the mean and variance of particle numbers changes as we move across spatial scales (by varying $m$) and to provide a systematic derivation of coarse-grained ($m>1$) models from the accurate ($m=1$) model. 

Sensible choices of $f^{(m)}$ require:
\begin{enumerate}
\item the volume exclusion function to be zero when the compartment is empty, $f^{(m)}(0)=0$;
\item the volume exclusion function to be unity when the compartment is at capacity, $f^{(m)}(m)=1$.
\end{enumerate}
One of the goals of this work is to elucidate functional forms for $f^{(m)}$ that give rise to behaviours that are conserved across spatial scales. We do this by considering equations for mean and variance of particle numbers.

% ----------------------------------------------------------------------------------------------------

\paragraph*{\bf Mean particle numbers.}

The evolution of mean particle number in the $j^{\text{th}}$ compartment when the transition probabilities are as in Eq.~\eqref{eq:TranisitionProbabilities} is given by the ordinary differential equation~\cite{footnote2}
\begin{eqnarray}
\label{eq:CollectingUpTermsForInitialMeanEquation}
\frac{\textrm{d} M^{(m)}_j}{\textrm{d}t}
&=&\frac{D}{m^2h^2}\left[-M^{(m)}_j+\left\langle n^{(m)}_{j}f^{(m)}\left(n^{(m)}_{j+1}\right)\right\rangle+M^{(m)}_{j-1}-\left\langle n^{(m)}_{j-1}f^{(m)}\left(n^{(m)}_j\right)\right\rangle\right]\nonumber\\
&&+\frac{D}{m^2h^2}\left[-M^{(m)}_j+\left\langle n^{(m)}_{j}f^{(m)}(n^{(m)}_{j-1})\right\rangle+M^{(m)}_{j+1}-\left\langle n^{(m)}_{j+1}f^{(m)}(n^{(m)}_j)\right\rangle\right],
\end{eqnarray}
for $2\leq{j}\leq{K-1}$, where $\langle\cdot\rangle$ denotes the expectation and $M^{(m)}_{j}=\left\langle{}n^{(m)}_{j}\right\rangle$. Similar expressions apply for the boundary compartments, $j=1,K$.

\paragraph*{The case $m=1$.} Here at most one particle can occupy each compartment, the conditions stated above are enough to fully define $f^{(1)}$, and we have~\cite{Liggett:1999:SIS} 
\begin{equation}
\label{equation:m1}
\dfrac{\textrm{d} M^{(1)}_j}{\textrm{d}t}=\frac{D}{h^2}\left(M^{(1)}_{j-1}-2M^{(1)}_j+M^{(1)}_{j+1}\right),
\end{equation}
for $2\leq{j}\leq{K-1}$. Similar expressions apply for the boundary compartments, $j=1,K$. Eq.~\eqref{equation:m1} is a semi-discrete diffusion equation and, in the limit $h\rightarrow0$, it gives rise to the diffusion equation with constant diffusion coefficient, $D$.

\paragraph*{The case $m=\infty$.} Letting $m\rightarrow\infty$ entails the limit of zero volume exclusion. To analyse the evolution of mean particle number we represent the compartment size as $\tilde{h}=L/K$ so that, similar to the $m=1$, case we have 
\begin{equation}
\label{equation:minfty}
\dfrac{\textrm{d} M^{(\infty)}_j}{\textrm{d}t}=\frac{D}{\tilde{h}^2}\left(M^{(\infty)}_{j-1}-2M^{(\infty)}_j+M^{(\infty)}_{j+1}\right),
\end{equation}
for $2\leq{j}\leq{K-1}$. Similar expressions apply for the boundary compartments, $j=1,K$. As for the $m=1$ case, Eq.~\eqref{equation:minfty} is a semi-discrete diffusion equation and, in the limit $\tilde{h}\rightarrow0$, it gives rise to the diffusion equation with constant diffusion coefficient, $D$.

\paragraph*{The case $1<m<\infty$.} To ensure the model is consistent across spatial scales, it is appropriate to confine choices for $f^{(m)}$ for $1<m<\infty$ to those that also give rise to a semi-discrete diffusion equation with constant diffusion coefficient for mean particle numbers. The only choice for $f^{(m)}$ is then
\begin{equation}
\label{eq:blockingprobability}
f^{(m)}\left(n_{j}^{(m)}\right)=\frac{n_{j}^{(m)}}{m},
\end{equation}
which gives, as anticipated,
\begin{equation}
\label{eq:MeanMasterEquation}
\dfrac{\textrm{d} M^{(m)}_j}{\textrm{d}t}=\frac{D}{m^2h^2}\left(M^{(m)}_{j-1}-2M^{(m)}_j+M^{(m)}_{j+1}\right),
\end{equation}
for $2\leq{j}\leq{K-1}$. Similar expressions apply for the boundary compartments, $j=1,K$. 

% ----------------------------------------------------------------------------------------------------

\paragraph*{\bf Variance of particle numbers.}

For the choice of volume exclusion function given in Eq.~\eqref{eq:blockingprobability} we can also obtain equations for the evolution of the variance of particle numbers:
\begin{eqnarray}
\label{eq:VarianceMasterEquation}
\dfrac{\mathrm{d}V_j^{(m)}}{\mathrm{d}t}
&=&
\frac{D}{m^2h^2}\left[\vphantom{\left(1-\frac{M_{j+1}^{(m)}}{m}\right)}2\left(\dfrac{m-1}{m}\right)V_{j,j-1}^{(m)}-4V_j^{(m)}+2\left(\dfrac{m-1}{m}\right)V_{j,j+1}^{(m)}\right]\nonumber\\
&&
+\frac{D}{m^2h^2}\left[M_{j-1}^{(m)}\left(1-\frac{M_{j}^{(m)}}{m}\right)+M_j^{(m)}\left(1-\frac{M_{j-1}^{(m)}}{m}\right)\right.\nonumber\\
&&\left.\qquad\qquad\qquad+M_j^{(m)}\left(1-\frac{M_{j+1}^{(m)}}{m}\right)+M_{j+1}^{(m)}\left(1-\frac{M_{j}^{(m)}}{m}\right)\right],\
\end{eqnarray}
for $2\leq{j}\leq{K-1}$, where $V^{(m)}_j$ is the variance of particle numbers in compartment $j$, and $V_{j,k}^{(m)}$ is the covariance of particle numbers in compartments $j$ and $k$:
\begin{eqnarray}
\label{eq:OffDiagCovarianceMasterEquation}
\dfrac{\mathrm{d}V_{j-1,j}^{(m)}}{\mathrm{d}t}
&=&
\frac{D}{m^2h^2}\left[\left(\dfrac{2}{m}-4\right)V_{j-1,j}^{(m)}+V_{j}^{(m)}+V_{j-1}^{(m)}+V_{j-2,j}^{(m)}+V_{j-1,j+1}^{(m)}\nonumber\right. \\
&&\qquad\qquad\left.-M_{j-1}^{(m)}\left(1-\dfrac{M_j^{(m)}}{m}\right)-M_j^{(m)}\left(1-\dfrac{M_{j-1}^{(m)}}{m}\right)\right];\\
\label{eq:DiffusionLikeCovarianceMasterEquation}
\dfrac{\mathrm{d}V_{j,k}^{(m)}}{\mathrm{d}t}
&=&\frac{D}{m^2h^2}\left[-4V_{j,k}^{(m)}+V_{j-1,k}^{(m)}+V_{j+1,k}^{(m)}+V_{j,k-1}^{(m)}+V_{j,k+1}^{(m)}\right] \nonumber \\ && \qquad\qquad \text{for}\quad 1<j<k-1<K, \quad 1<k+1<j<K.
\end{eqnarray}
Similar expressions can be found for the boundary compartments, $j=1,K$. 

% ----------------------------------------------------------------------------------------------------

\section{Consistency of the choice of volume exclusion function}\label{sec:justification}

We consider the $m=1$ case to represent the most `accurate' model of volume exclusion effects for an on-lattice model of diffusion as it implies, simply, that one particle cannot overlap with another. When coarse-graining this model, to consider random walk models with compartment capacities $m>1$, our aim is that the mean and variance of particle numbers in each compartment are conserved. In what follows, we will sum the mean and variance of compartment occupancy of the accurate, $m=1$, model over groups of $m$ contiguous compartments to explore how accurate we can expect the coarse-grained model to be.

To this end, we will define 
\begin{equation}
S_{j}^{(m)}(t)=\sum\limits_{i=(j-1)m+1}^{jm}n_{i}^{(1)}(t),
\end{equation}
with $\mu_{j}^{(m)}(t)$ the mean of $S_{j}^{(m)}(t)$, and $v_{j}^{(m)}(t)$ its variance. We wish to establish the relationship between: (i) $\mu_{j}^{(m)}(t)$ and $M_{j}^{(m)}(t)$; and (ii) $v_{j}^{(m)}(t)$ and $V_{j}^{(m)}(t)$.

% ----------------------------------------------------------------------------------------------------

\paragraph*{\bf Steady state values.}

Eqs.~\eqref{eq:MeanMasterEquation}-\eqref{eq:DiffusionLikeCovarianceMasterEquation}, together with the additional constraint that the sum of all variance and covariance terms must be zero (since $N$ is constant), gives the steady states
\begin{eqnarray}
 \hat{M}_{i}^{(m)}&=&\dfrac{N}{K},\\
 \hat{V}_{i}^{(m)}&=&\dfrac{N(K-1)}{K\left(K-\frac{1}{m}\right)}\left(1-\dfrac{N}{mK}\right),\\
 \hat{V}_{i,j}^{(m)}&=&\dfrac{-N}{K\left(K-\frac{1}{m}\right)}\left(1-\dfrac{N}{mK}\right),\;\; i\neq j,
\end{eqnarray}
where $1\le i,j\le K$ and circumflexes are used to denote steady state values. It is then simple to check that
\begin{eqnarray}
\hat{\mu}_{j}^{(m)}&=&\hat{M}_{j}^{(m)},\\
\hat{v}_{j}^{(m)}&=&\hat{V}_{j}^{(m)},
\end{eqnarray}
\textit{i.e.} the steady state means and variances are conserved through the process of coarse-graining.

% ----------------------------------------------------------------------------------------------------

\paragraph*{\bf Time evolution.}

To obtain an expression for the evolution of $\mu_{j}^{(m)}$, we note 
\begin{eqnarray}
\label{eq:SumOfExcludingMeanEquations}
\dfrac{\mathrm{d}\mu_{j}^{(m)}}{\mathrm{d}t}
&=&\sum\limits_{i=(j-1)m+1}^{jm}\dfrac{\mathrm{d}M_{i}^{(1)}}{\mathrm{d}t}\nonumber\\
&=&\dfrac{D}{h^2}\sum_{i=(j-1)m+1}^{jm}\left(M_{i-1}^{(1)}-2M_{i}^{(1)}+M_{i+1}^{(1)}\right)\nonumber\\
&=&\dfrac{D}{h^2}\left(M_{(j-1)m}^{(1)}-M_{(j-1)m+1}^{(1)}-M_{jm}^{(1)}+M_{jm+1}^{(1)}\right).\nonumber\\
\end{eqnarray}
We compare this to the coarse-grained model, Eq.~\eqref{eq:MeanMasterEquation}, which we re-state here for convenience:
\begin{equation}
\nonumber
\dfrac{\mathrm{d}M_{j}^{(m)}}{\mathrm{d}t}=\dfrac{D}{m^2h^{2}}\left(M_{j-1}^{(m)}-2M_{j}^{(m)}+M_{j+1}^{(m)}\right).
\end{equation}
To relate Eqs.~\eqref{eq:MeanMasterEquation} and~\eqref{eq:SumOfExcludingMeanEquations}, and understand when we expect the coarse-grained model to replicate the dynamics of the accurate, $m=1$, model, we need to establish a relationship between the $M_j^{(1)}$ ($j=1,\ldots,N$) and the $M_j^{(m)}$ ($j=1,\ldots,K$). 

A natural choice for the coarse-graining would be to assume $M_i^{(1)}\approx{}M_{j}^{(m)}/m$ for $i=(j-1)m+1,\ldots,jm$. However, Eq.~\eqref{eq:SumOfExcludingMeanEquations} would then give rise to a diffusion equation with constant diffusion coefficient $m$ times larger than expected in the limit $h\rightarrow0$. This means that our coarse-grained model does \emph{not} require the stringent condition that particles in a compartment of size $m$ are uniformly distributed throughout that compartment.

\begin{figure}[h]
\includegraphics{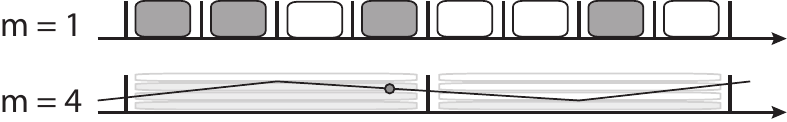}
\centering
\caption{Schematic of the interpolation process for $m=4$.}
\label{fig:interpolation}
\end{figure}

Instead, consistency between the accurate ($m=1$) and the coarse-grained ($m>1$) models arises from assuming particles in the $m>1$ compartments are distributed, on average, according to a linear interpolation between $m>1$ neighbouring compartments rather than being uniformly distributed throughout the compartment:
\begin{eqnarray}
M_{(j-1)m}^{(1)}&=&\frac{1}{m}\left(\dfrac{1}{2}\dfrac{m+1}{m}\mu_{j-1}^{(m)}+\dfrac{1}{2}\dfrac{m-1}{m}\mu_{j}^{(m)}\right),\nonumber\\
M_{(j-1)m+1}^{(1)}&=&\frac{1}{m}\left(\dfrac{1}{2}\dfrac{m-1}{m}\mu_{j-1}^{(m)}+\dfrac{1}{2}\dfrac{m+1}{m}\mu_{j}^{(m)}\right),\nonumber\\
M_{jm}^{(1)}&=&\frac{1}{m}\left(\dfrac{1}{2}\dfrac{m+1}{m}\mu_{j}^{(m)}+\dfrac{1}{2}\dfrac{m-1}{m}\mu_{j+1}^{(m)}\right),\nonumber\\
M_{jm+1}^{(1)}&=&\frac{1}{m}\left(\dfrac{1}{2}\dfrac{m-1}{m}\mu_{j}^{(m)}+\dfrac{1}{2}\dfrac{m+1}{m}\mu_{j+1}^{(m)}\right),\nonumber \\
&&
\end{eqnarray}
as shown in Fig.~\ref{fig:interpolation}. As a result we have
\begin{equation}
\dfrac{\mathrm{d}\mu_{j}^{(m)}}{\mathrm{d}t}=\dfrac{D}{m^2h^{2}}\left(\mu_{j-1}^{(m)}-2\mu_{j}^{(m)}+\mu_{j+1}^{(m)}\right),
\end{equation}
and evolution of the mean particle numbers in the coarse-grained system matches that of the accurate model.

% ----------------------------------------------------------------------------------------------------

% ----------------------------------------------------------------------------------------------------

The entries of the covariances matrix $\left\{V_{i,j}^{(1)}\right\}$ cannot be interpolated in the same way, since its entries are positive on the diagonal and negative everywhere else. However, we can use similar reasoning to argue that the variances $v_{j}^{(m)}$ and $V_{j}^{(m)}$ will also match, as presented in the Supplemental Material.

% ----------------------------------------------------------------------------------------------------

\section{Numerical investigations}\label{sec:NumericalExamples}

We now present numerical results to corroborate our findings. We consider the domain $x\in[0,1]$ with $h=1/128$ and $D=10^3$. The initial condition is $n_j^{(1)}(0)=1$, for $j=1,\ldots,16$, and $n_j^{(1)}(0)=0$ otherwise, and attempts by particles to jump left out of compartment 1 or right out of compartment 128 are aborted. 

\begin{figure}
\includegraphics{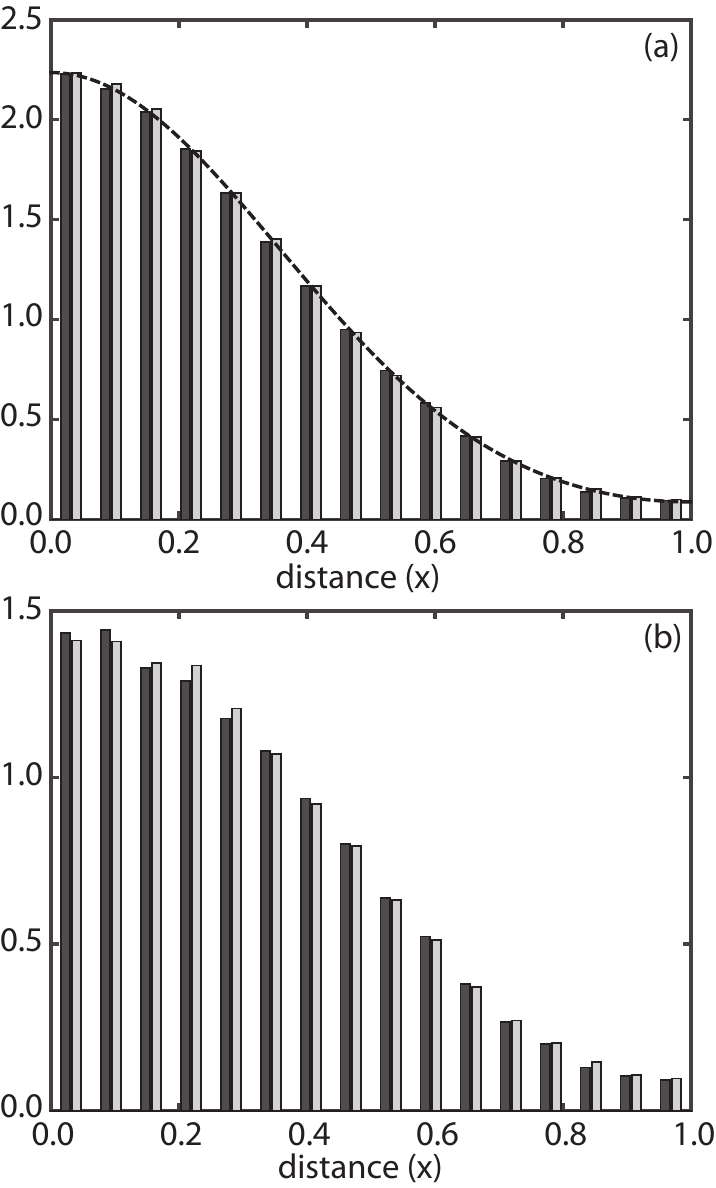}
\caption{Comparing the mean (a) and variance (b) of particle numbers for the $m=1$ and $m=8$ cases at $t=1$, with $K=128/m$. Dark grey bars: results from simulation of the random walk model with $m=1$, light grey bars: results from numerical solution of Eq.~\eqref{eq:MeanMasterEquation}/Eq.~\eqref{eq:VarianceMasterEquation} with $m=8$. Black dashed line in (a): solution of the limiting diffusion equation.}
\label{fig:meansandvariances}
\end{figure}

We compare results from 5000 realisations of the random walk model with $m=1$ with 5000 realisations of the same model with $m=8$ in Fig.~\ref{fig:meansandvariances}. The mean values predicted using both the $m=8$ and PDE models are in excellent agreement with those predicted from the accurate $m=1$ model. In addition, we see good agreement between the variances of the $m=1$ and $m=8$ models~\cite{footnote3}. Finally, we note that an additional advantage of the coarse-grained model is that generating realisations of the discrete random walk model with $m>1$ can be achieved in $1/m^2$ of the time required by the $m=1$ case, since the jump rates will be $m^2$ times smaller and so far fewer jumps will need to be simulated.

% ----------------------------------------------------------------------------------------------------

% ----------------------------------------------------------------------------------------------------

\begin{figure}
\includegraphics{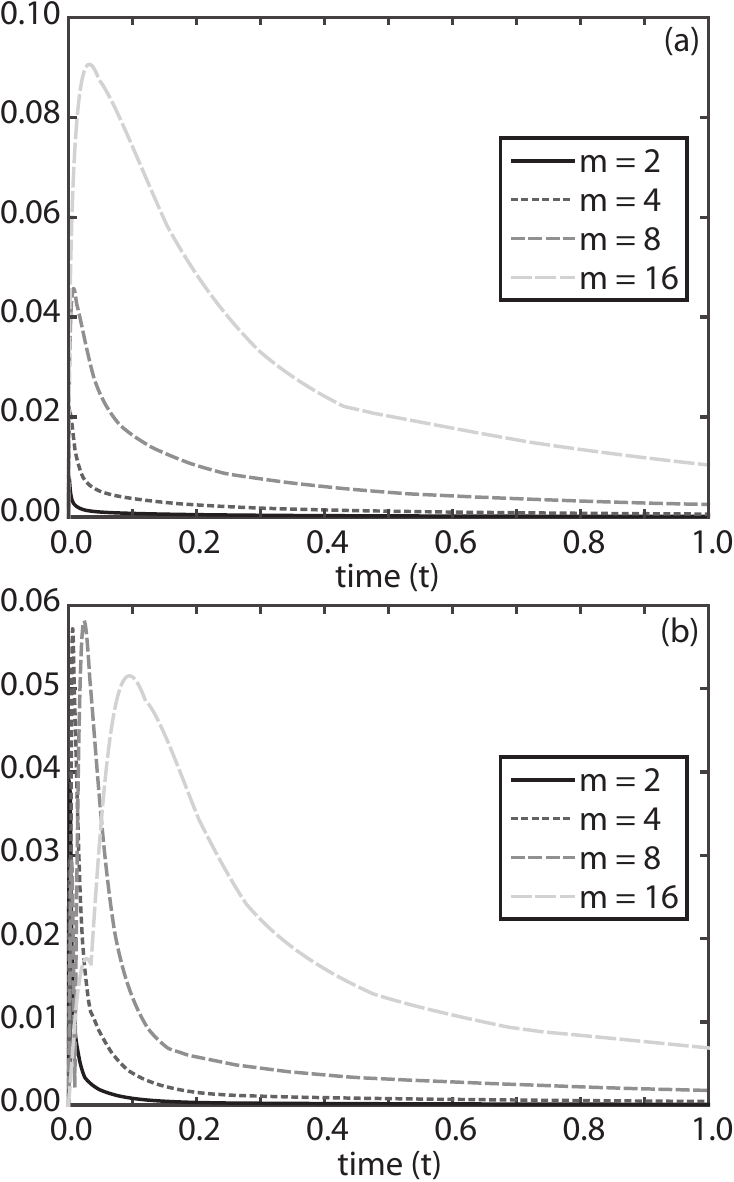}
\caption{HDEs for the means (a) and variances (b) in particle numbers as $m$ is varied~\cite{footnote3}.}
\label{fig:HDEs}
\end{figure}

To compute the error in the coarse-graining process, we solved the ODEs for both the means and variances, Eqs.~\eqref{eq:MeanMasterEquation} and~\eqref{eq:VarianceMasterEquation}, over a range of values of $m$~\cite{footnote4}. Results for $m>1$ were compared against results with $m=1$ using the histogram distance error (HDE) metric~\cite{Cao:2006:ALA}:
\begin{equation}
\mathrm{HDE}=\frac{1}{2}\sum\limits_{k=1}^{K}|e_k-p_k|,
\end{equation}
where $e_k$ is the normalised value of the $k^\text{th}$ aggregated compartment of the $m=1$ model and $p_k$ is the normalised value of the $k^\text{th}$ compartment of the $m>1$ model. Figure~\ref{fig:HDEs} demonstrates evolution of the error between the models with different values of $m$ between $t=0$ and $t=1$. The HDE remains low in all cases observed, even though the initial condition does not satisfy the requirement that the densities in the $m=1$ case can be interpolated onto the coarse ($m>1$) lattice.

%% ----------------------------------------------------------------------------------------------------

% ----------------------------------------------------------------------------------------------------

\section{Conclusions}\label{sec:Conclusions}

In this work we have used an on-lattice random walk approach to reconcile models of diffusive transport that incorporate the effects of excluded volume across spatial scales. Our work demonstrates that coarse-grained models incorporating simplified descriptions of excluded volume can be used in many circumstances, and these simplified models engender significantly lower computational costs than their accurate counterparts. These computational savings are expected to be especially valuable for models in two or three spatial dimensions. However, care must be taken in pushing the coarse-graining process too far. For example, there is a delicate trade off between the initial conditions of the model and the size of $m$. Future work will be directed towards hybrid approaches, in which a detailed description of the spatial dynamics can be retained where necessary, and the computational savings associated with the coarse-grained model taken advantage of where possible.

\medskip

% ----------------------------------------------------------------------------------------------------

\begin{acknowledgments}
PRT gratefully acknowledges an EPSRC studentship through the University of Oxford's Systems Biology Doctoral Training Centre. MJS acknowledges support from the Australian Research Council (FT130100148).
\end{acknowledgments}

% ----------------------------------------------------------------------------------------------------

\appendix

% ----------------------------------------------------------------------------------------------------

\section{Derivation of mean and variance equations}\label{Appendix:DerivationOfMeanEquations}

Recall that we have defined a spatial lattice composed of $K$ compartments. The distribution of particles over this domain is given by the vector $\textbf{n}(t)=[n_1(t),n_2(t),...,n_K(t)]$, where $n_i(t)$ denotes the number of particles in the $i^{\mathrm{th}}$ compartment at time $t$. Each lattice compartment has capacity $m$, but in the interests of clarity we will omit the superscript $(m)$ from the variables. Particles in compartment $i$ may \textit{attempt} to jump out to compartments $i-1$ or $i+1$: jumps in each direction are equally likely, and occur with rate per unit time of $d$.

We define two operators, $J_{i}^{+}:\mathbb{R}^{K}\rightarrow\mathbb{R}^{K}$, for $i=1,...,K-1$, and 
$J_{i}^{-}:\mathbb{R}^{K}\rightarrow\mathbb{R}^{K}$, for $i=2,...,K$, as
\begin{eqnarray}
&J_{i}^{+}:[n_1,...,n_i,...,n_K] \rightarrow [n_1,...,n_{i-2},n_{i-1},n_{i}+1,n_{i+1}-1,n_{i+2},...,n_K], \\
&J_{i}^{-}:[n_1,...,n_i,...,n_K] \rightarrow [n_1,...,n_{i-2},n_{i-1}-1,n_{i}+1,n_{i+1},n_{i+2},...,n_K].
\end{eqnarray}
Both operators move a particle into compartment $i$, taken from the compartment to the right or left, respectively. We assume that attempted jumps into some compartment $j$ fail with probability $f(n_j)$ due to exclusion effects. We can then write the probability master equation as
\begin{eqnarray}
\label{eq:lastprobequationbeforemeans}
\frac{\textrm{d} \mathrm{Pr}(\textbf{n},t)}{\textrm{d} t}&=&\sum\limits_{i=1}^{K-1}d\left\{(n_i+1)\left[1-f(n_{i+1}-1)\right]\mathrm{Pr}(J_i^+\textbf{n},t)-n_i\left[1-f(n_{i+1})\right]\mathrm{Pr}(\textbf{n},t)\right\} \nonumber\\
&&+\sum\limits_{i=2}^{K}d\left\{(n_i+1)\left[1-f(n_{i-1}-1)\right]\mathrm{Pr}(J_i^-\textbf{n},t)-n_i\left[1-f(n_{i-1})\right]\mathrm{Pr}(\textbf{n},t)\right\}.
\end{eqnarray}
Define mean vector $\textbf{M}=[M_1,...,M_K]$, where
\begin{equation}
M_j=\sum\limits_{n_1=1}^{N}\sum\limits_{n_2=1}^{N}\dots\sum\limits_{n_K=1}^{N}n_j\mathrm{Pr}(\textbf{n},t):=\sum\limits_{n_1,n_2,...,n_K=0}^{\mathcal{N}}n_j\mathrm{Pr}(\textbf{n},t).
\end{equation}
Multiplying Eq. \eqref{eq:lastprobequationbeforemeans} by $n_j$ and summing over all possible values that the vector $\textbf{n}(t)$ can take we have
\begin{eqnarray}
\label{eq:PME_start}
\frac{\textrm{d} M_j}{\textrm{d} t}&=&\sum\limits_{n_1,n_2,...,n_K=0}^{\mathcal{N}}n_j\left(\sum\limits_{i=1}^{K-1}d\left\{(n_i+1)\left[1-f(n_{i+1}-1)\right]\mathrm{Pr}(J_i^+\textbf{n},t)-n_i\left[1-f(n_{i+1})\right]\mathrm{Pr}(\textbf{n},t)\right\}\right. \nonumber\\
&&\qquad\qquad\left.+\sum\limits_{i=2}^{K}d\left\{(n_i+1)\left[1-f(n_{i-1}-1)\right]\mathrm{Pr}(J_i^-\textbf{n},t)-n_i\left[1-f(n_{i-1})\right]\mathrm{Pr}(\textbf{n},t)\right\}\right).
\end{eqnarray}
We begin by considering only the first term of this expression, with $Pr(J_i^{+}\textbf{n},t)$ expanded explicitly to give
\begin{equation}
 \sum\limits_{n_1,n_2,...,n_K=0}^{\mathcal{N}}n_j\sum\limits_{i=1}^{K-1}d(n_i+1)\left[1-f(n_{i+1}-1)\right]Pr(n_1,n_2,...,n_i+1,n_{i+1}-1,...n_K,t).
\end{equation}
When $i\neq j,j-1$ each term of this expression reduces to
\begin{equation}
d\langle n_jn_i\rangle-d\langle n_jn_if(n_{i+1})\rangle,
\end{equation}
where $\langle n_jn_i\rangle$ indicates the mean of the product. When $i=j$,
\begin{eqnarray}
&\sum\limits_{n_1,n_2,...,n_K=0}^{\mathcal{N}}n_jd(n_j+1)\left[1-f(n_{j+1}-1)\right]Pr(n_1,n_2,...,n_j+1,n_{j+1}-1,...n_K,t), \nonumber\\
=&\sum\limits_{n_1,n_2,...,n_K=0}^{\mathcal{N}}d(n_j+1)^2\left[1-f(n_{j+1}-1)\right]Pr(n_1,n_2,...,n_j+1,n_{j+1}-1,...n_K,t) \nonumber\\
&-d(n_j+1)\left[1-f(n_{j+1}-1)\right]Pr(n_1,n_2,...,n_j+1,n_{j+1}-1,...n_K,t),\nonumber\\
=&d[\langle n_jn_j\rangle-\langle n_jn_jf(n_{j+1})\rangle-M_j+\langle n_jf(n_{j+1})\rangle].
\end{eqnarray}
Similarly for $i=j-1$,
\begin{eqnarray}
&\sum\limits_{n_1,n_2,...,n_K=0}^{\mathcal{N}}dn_j(n_{j-1}+1)\left[1-f(n_{j}-1)\right]Pr(n_1,n_2,...,n_{j-1}+1,n_{j}-1,...n_K,t), \nonumber\\
=&d[\langle n_{j-1}n_j\rangle-\langle n_{j-1}n_jf(n_j)\rangle+M_{j-1}-\langle n_{j-1}f(n_j)\rangle].
\end{eqnarray}
We then consider the second term,
\begin{equation}
-\sum\limits_{n_1,n_2,...,n_K=0}^{\mathcal{N}}n_j\sum\limits_{i=1}^{K-1}dn_i\left[1-f(n_{i+1})\right]Pr(n_1,n_2,...,n_i,...,n_K,t),
\end{equation}
which evaluates to
\begin{equation}
-\sum\limits_{i=1}^{K-1} d\left[\langle n_in_j\rangle-\langle n_if(n_{i+1})n_j\rangle\right].
\end{equation}
When combined with the expressions derived from the first term these give us
\begin{equation}\label{eq:firsthalfofmeanPME}
 -M_j+\langle n_{j}f(n_{j+1})\rangle+M_{j-1}-\langle n_{j-1}f(n_j) \rangle.
\end{equation}
Applying the same approach to the third and fourth terms, we obtain
\begin{equation}
 -M_j+\langle n_{j}f(n_{j-1})\rangle+M_{j+1}-\langle n_{j+1}f(n_j) \rangle.
\end{equation}
Adding this expression to (\ref{eq:firsthalfofmeanPME}), we arrive at
\begin{eqnarray}
\frac{\textrm{d} M_j}{\textrm{d}t}&=&d\left[-M_j+\langle n_{j}f(n_{j+1})\rangle+M_{j-1}-\langle n_{j-1}f(n_j) \rangle\right]\nonumber\\
&&+d\left[-M_j+\langle n_{j}f(n_{j-1})\rangle+M_{j+1}-\langle n_{j+1}f(n_j) \rangle\right].
\end{eqnarray}
Similar reasoning can be applied to derive equations for $M_1$ and $M_K$ and this approach can also be used to derive the variance and covariance equations, where we use
\begin{equation}
V_j(t)=\sum\limits_{n_1,n_2,...,n_K=0}^{\mathcal{N}}\left[n_{j}^{2}\mathrm{Pr}(\textbf{n},t)\right]-M_{j}^{2}(t).
\end{equation}
It can be shown that
\begin{eqnarray}
\dfrac{\mathrm{d}}{\mathrm{d}t}\sum\limits_{n_1,n_2,...,n_K=0}^{\mathcal{N}}n_{i}^{2}P(\textbf{n})&=&d\left\{2\left(1-\dfrac{1}{m}\right)\langle n_{i-1}n_{i}\rangle-4\langle n_{i}^2\rangle+2\left(1-\dfrac{1}{m}\right)\langle n_{i}n_{i+1}\rangle+M_{i-1}+2M_i+M_{i+1}\right\}.\nonumber\\
\end{eqnarray}
By using $\langle n_in_j\rangle=V_{i,j}+M_iM_j$, we can then write
\begin{eqnarray}
\dfrac{\mathrm{d}V_i}{\mathrm{d}t}&=&\dfrac{\mathrm{d}}{\mathrm{d}t}\sum\limits_{n_1,n_2,...,n_K=0}^{\mathcal{N}}n_i^2P(\textbf{n})-\dfrac{\mathrm{d}}{\mathrm{d}t}M_i^2\nonumber\\
&=&\dfrac{\mathrm{d}}{\mathrm{d}t}\sum\limits_{\mathcal{N}}n_i^2P(\textbf{n})-2d\left(M_{i-1}M_{i}-2M_{i}^2+M_{i}M_{i+1}\right),\nonumber\\
&=&d\left[2\left(\dfrac{m-1}{m}\right)V_{i,i-1}-4V_i+2\left(\dfrac{m-1}{m}\right)V_{i,i+1} \right.\nonumber\\
&&\qquad+M_{i-1}\left(1-\frac{M_{i}}{m}\right)+M_i\left(1-\frac{M_{i-1}}{m}\right)\nonumber\\
&&\qquad\left.+M_i\left(1-\frac{M_{i+1}}{m}\right)+M_{i+1}\left(1-\frac{M_{i}}{m}\right)\right],
\end{eqnarray}
with similar results obtainable for the covariance terms, and for $i=1$ and $i=K$.

% ----------------------------------------------------------------------------------------------------

\section{Matching variance equations}\label{Appendix:VarianceComparison}

We begin by noting that Eq. (8) of the main text is essentially a semi-discrete two-dimensional diffusion equation and has no dependence on $m$, except in determining the jump rate, so it is clear that, away from the diagonal of the covariance matrix, the covariances of the $m>1$ model will evolve similarly to those of the summed $m=1$ model.

To consider the behaviour around the variance values, and adjacent covariances, we begin by noting that with $m=1$, Eq. (7) of the main text becomes
\begin{eqnarray}
\dfrac{\mathrm{d}V_i}{\mathrm{d}t}&=&\dfrac{D}{h^2}\left[-4V_i+M_{i-1}\left(1-M_{i}\right)+M_{i}\left(1-M_{i-1}\right) \right.\nonumber\\
&&\qquad\left.+M_{i}\left(1-M_{i+1}\right)+M_{i+1}\left(1-M_{i}\right)  \right],
\end{eqnarray}
and
\begin{eqnarray}
\dfrac{\mathrm{d}V_{i-1,i}}{\mathrm{d}t}&=&\dfrac{D}{h^2}\left[-2V_{i-1,i}+V_{i}+V_{i-1}+V_{i-2,i}+V_{i-1,i+1} \right.\nonumber\\
&&\qquad\left.+M_{i-1}\left(1-M_{i}\right)+M_{i}\left(1-M_{i-1}\right)\right].
\end{eqnarray}
By the standard result for the variance of the sum of random variables, we find
\begin{eqnarray}
\dfrac{\mathrm{d}v_j^{(m)}}{\mathrm{d}t}&=&\sum\limits_{k=(j-1)m+1}^{jm}\left(\dfrac{\mathrm{d}V_{k}^{(1)}}{\mathrm{d}t}\right)+2\sum\limits_{p=(j-1)m+1}^{jm-1}\sum\limits_{q=p+1}^{jm}\left(\dfrac{\mathrm{d}V_{p,q}^{(1)}}{\mathrm{d}t}\right).
\end{eqnarray}
We evaluate each of the terms on the right-hand side to obtain
\begin{eqnarray}
\frac{\textrm{d}v_j^{(m)}}{\textrm{d}t}&=&\dfrac{D}{h^2}
\left[\sum\limits_{p=(j-1)m+1}^{jm}\left(V_{p,(j-1)m+1}^{(1)}+V_{p,jm}^{(1)}\right)\right.-\sum\limits_{p=(j-1)m+2}^{jm}\left(V_{p,(j-1)m}^{(1)}+V_{p,jm+1}^{(1)}\right)\nonumber\\
&&\qquad\qquad+\sum\limits_{q=(j-1)m+1}^{jm}\left(V_{(j-1)m+1,q}^{(1)}+V_{jm,q}^{(1)}\right)-\sum\limits_{q=(j-1)m+2}^{jm}\left(V_{(j-1)m,q}^{(1)}+V_{jm+1,q}^{(1)}\right)\nonumber\\
&&\qquad\qquad+M_{(j-1)m}^{(1)}\left(1-\dfrac{M_{(j-1)m+1}^{(1)}}{m}\right)+M_{(j-1)m+1}^{(1)}\left(1-\dfrac{M_{(j-1)m}^{(1)}}{m}\right)\nonumber\\
&&\qquad\qquad+M_{jm}^{(1)}\left(1-\dfrac{M_{jm+1}^{(1)}}{m}\right)\left.+M_{jm+1}^{(1)}\left(1-\dfrac{M_{jm}^{(1)}}{m}\right)\vphantom{\sum\limits_{p=(j-1)m+1}^{jm}}\right].
\end{eqnarray}
We may draw an analogy to the mean equations, when after summation only those squares on the borders of the notional coarse grid compartment make a contribution to the dynamics; we wish to show that $v_j^{(m)}$ evolves in the same way as $V_j^{(m)}$ does in Eq. (7),
\begin{eqnarray}
 \dfrac{\mathrm{d}V_i}{\mathrm{d}t}&=&\dfrac{d}{m^2}\left[2\left(\dfrac{m-1}{m}\right)V_{i,i-1}-4V_i+2\left(\dfrac{m-1}{m}\right)V_{i,i+1} \right.\nonumber\\
&&\qquad\qquad+M_{i-1}\left(1-\dfrac{M_{i}}{m}\right)+M_i\left(1-\dfrac{M_{i-1}}{m}\right)\nonumber\\
&&\qquad\qquad\left.+M_i\left(1-\dfrac{M_{i+1}}{m}\right)+M_{i+1}\left(1-\dfrac{M_{i}}{m}\right)\right].
\end{eqnarray}

% ----------------------------------------------------------------------------------------------------

\begin{figure}
\includegraphics[width=0.7\textwidth]{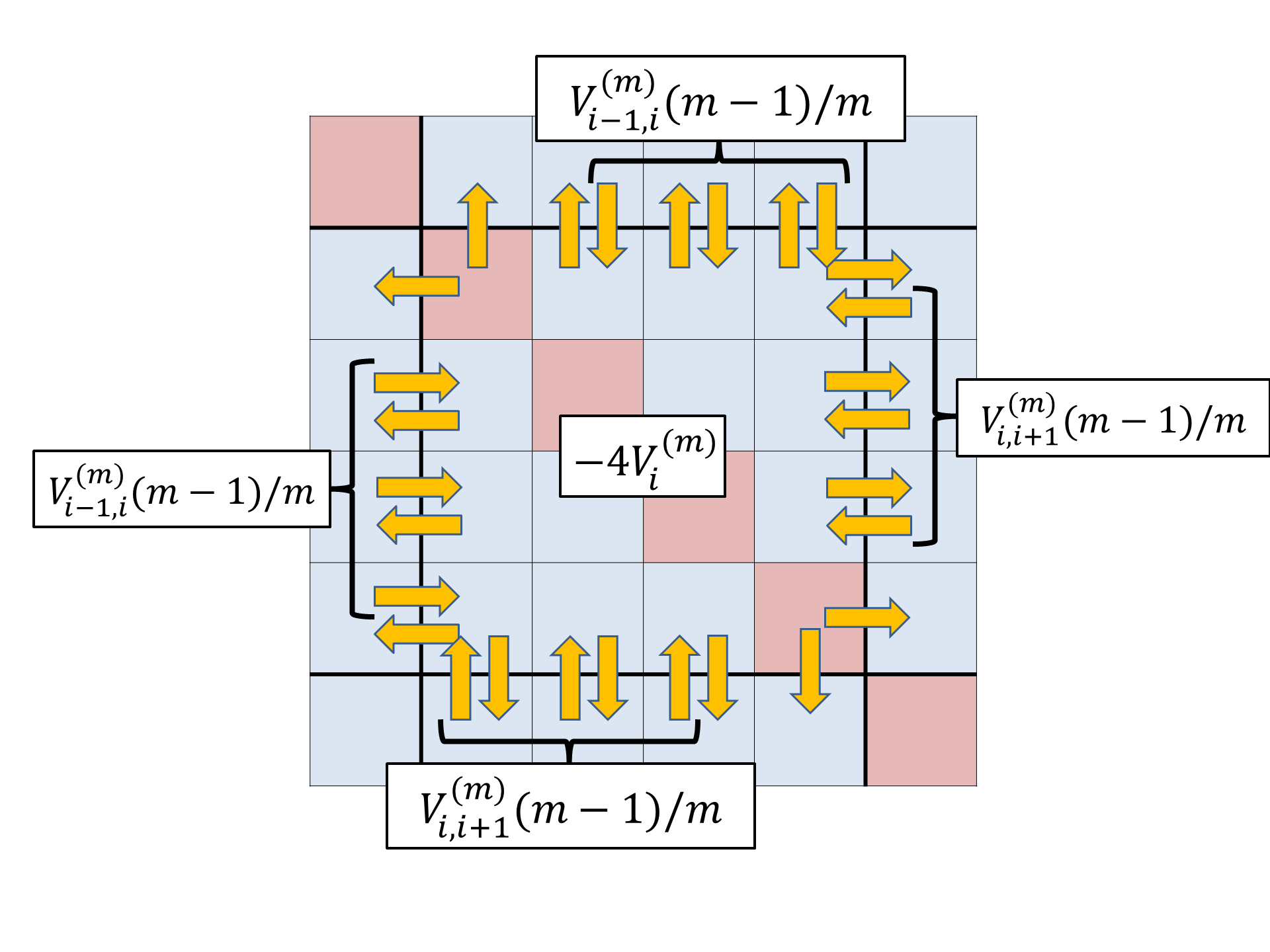}
\centering
\caption{The red compartments represent the variances of the fully-excluding system (i.e. the central diagonal of the covariance matrix), while the blue compartments represent the covariances. It can be seen that particles are leaving the central compartment at every edge (accounting for the $-4V^{(m)}_i$ term in the coarse grid model), and are entering from the surrounding compartments at each edge except for the two in the top-left corner and the two in the bottom-right corner (those adjacent to the fine grid variance terms). This explains why the input terms are $V^{(m)}_{i-1,i}(m-1)/m$ and $V^{(m)}_{i,i+1}(m-1)/m$, respectively.}
\label{fig:VarianceScalingDiagram}
\end{figure}

% ----------------------------------------------------------------------------------------------------

The contributions of the variance terms are illustrated in Figure \ref{fig:VarianceScalingDiagram}. It can be seen that they represent a simple diffusion pattern in two dimensions similar to the expressions seen in the mean equations. The only deviation from this is seen in the top-left and bottom-right corners, where there is outflow to the external cells but no inflow from them. Since the affected region on each edge is one compartment-length out of $m$, it seems intuitive that this would correspond to the $m>1$ model, where the input terms from the surrounding compartments all scale with $(m-1)/m$. A similar argument can be made for the $v_{j-1,j}^{(m)}$ terms, where outflow scales with $(4-2/m)$. We also note that in our expression for $v_j^{(m)}$, the contributions from mean terms simply represent the propensity for a move occuring, just as they do in the partially-excluding case. On this basis, and considering the strong agreements observed in Section IV of the main text, it seems reasonable to expect 
the variance of the total number of particles across $m$ fully-excluding compartments to match that of a partially-excluding compartment with capacity $m$.

% ----------------------------------------------------------------------------------------------------

% ----------------------------------------------------------------------------------------------------


\begin{thebibliography}{}
\bibitem{footnote1} Note that the zero flux boundary conditions effectively entail that jumps left from compartment $1$ or right from compartment $K$ have `zero rate'.
\bibitem{footnote3} Realisations of the random walk model were generated using the Gillespie algorithm, and the limiting diffusion PDE using the Matlab routine \textsf{pdepe} with $\delta{x}=5\times10^{-4}$.
\end{thebibliography}
\end{document}